\documentstyle{article}
\author{Piret Kuusk and Madis K\~oiv\\
Institute of Physics, University of Tartu,
Riia 142, 51014 Tartu, Estonia;\\ 
piret@fi.tartu.ee}
\title{Measurement of time in nonrelativistic 
quantum and classical mechanics}  
\date{\mbox{}}
\begin{document}
\maketitle

\begin{abstract}
Possible theoretical frameworks for measurement    
of (arrival) time in the 
nonrelativistic quantum mechanics are reviewed. It is argued that 
the ambiguity between
indirect measurements by a suitably introduced time operator and 
direct measurements by a physical clock particle  has a
counterpart   in the corresponding classical framework of measurement
of the Newtonian time based on the  Hamiltonian  mechanics.

\end{abstract}


\section{Introduction}

The problem of time in the nonrelativistic quantum mechanics 
consists in the following dichotomy: a measurement of time can be
described  as a statistical distribution of measurement
outcomes given by a suitable time operator (or by the corresponding
spectral measure) canonically conjugate
to the energy operator, or the time flow can be 
visualized by the change in the position (or in the momentum) 
observable of a physical clock particle
which interacts  with other parts of the system under consideration
and the total wave function must be 
determined by a quantum evolution equation 
(e.g. by a Schr\"odinger equation). 
It is not clear if these two possibilities
can formally be identified. The problem was 
discussed in the classical paper by Aharonov and Bohm \cite{AB}, but 
has recently gained considerable new development  \cite{physrep, book}.
Our aim here is to clarify some aspects of the quantum theory
by analyzing more in detail the corresponding classical 
theory of measurement of time.        

In fact, the above-mentioned dichotomy exists already in the
classical theory:  time can be understood as
a parameter along a classical trajectory
or as a record of  change in the spatial position
of a special clock particle.
Although these two possibilities are classically  equivalent, 
their precise mathematical description in the framework of 
the Hamiltonian formalism is not identical and upon canonical 
quantization they can give rise to quantum theories of time which need not be
equivalent. The key to the understanding of the problem of time
in quantum mechanics may lie in the classical theory; so  we can
conclude together with   L\'evy-Leblond 
\cite{LL}: {\it nobody has ever constructed a complete "classical theory of 
measurement"} and
{\it "To the initial question "where is the problem (...)?",
I would therefore venture the paradoxical answer: "in}   classical
{\it theory"...} 

\section{The quantum theory}

\subsection{Operators of position and time}

Nonrelativistic quantum mechanics is a theory of (quantized) 
matter in the background of a nonrelativistic spacetime with
Newtonian space coordinates ($x^i$) and Newtonian time
coordinate $t$ \cite{Holevo, Ludwig, BGLOperat}.
The symmetry group of the background coordinates is the
Galilei group    
$$
(x^i)^\prime = x^i - \xi^i -V^i t \,,  \qquad
t^\prime = t - \tau \,.
$$
Here $\xi^i$, $V^i$,  $\tau$ are the group parameters and we haven't included 
the group of spatial rotations. 

The Galilei group has a (projective irreducible) representation acting
on  density matrices $\rho$ (states of quantized matter). 
For simplicity, let us consider  one-dimensional space, then the 
two-dimensional subgroup of the Galilei group 
with parameters ($\xi,V$) can be represented
by unitary operators $\exp(imV Q)$, $ \exp(-i\xi P)$, $m$ =const.  
Self-adjoint operators $Q$ and $P$ satisfy 
the canonical commutation  relations  
$$
[Q, P] = i I,
$$
where $I$ is the unit operator. These operators can be given in terms of
the corresponding spectral measures $E^Q(x)$, $E^P(p)$:
$$
Q =\int_R x E^Q(dx), \qquad P = m \int_R  p E^P(dp) .
$$
They can be interpreted as the position and the momentum operators
if they satisfy covariance conditions under the action of the Galilei group, 
e.g.
\begin{equation}
e^{i\xi P} E^Q(x) e^{-i\xi P} = E^Q(x-\xi) ,
 \qquad \xi \in (-\infty, +\infty).                                  
\label{spacecov}
\end{equation} 
This means  that if the device for registration of the position 
(spectral measure $E^Q$) undergoes 
a spatial shift in the amount of $\xi$
then  measured values (spectrum) change by $(-\xi)$. 

If the same scheme is applied in the case of 
 canonically conjugate operators of time $T $ and energy $H$,
$[T, H] =- i I$,
then  operator $T$,
for its interpretation as an operator which establishes a relation
between a quantum state and the Newtonian time, must satisfy the
covariance property
\begin{equation}
e^{-i\tau H} E^T(t) e^{i\tau H} = E^T(t-\tau), 
\qquad \tau \in (-\infty, + \infty).
                         \label{timecov}
\end{equation}
It can be proved that if the time operator is self-adjoint, then this 
condition implies an unbounded continous energy spectrum (Pauli's theorem)
\cite{Holevo}.  It follows that
in the cases of bounded, semi-bounded or discrete energy spectra
the covariant time operator cannot be self-adjoint, i.e its 
spectral measure $E^T(t) $ cannot be given by projection operators. 
However, it can be given as a positive operator valued measure (POVM)
\cite{Holevo, BGLOperat}. There can also be non-covariant self-adjoint 
operators canonically conjugate to the energy operator \cite{Galapon}.
But if covariance condition (\ref{timecov}) is not satisfied,
their interpretation as  time operators is problematic.      
 
If the notion of time in the quantum mechanics is understood  as 
a record of a quantum measurement, then it can depend on the measurement
scheme and  need not be unique.  Indeed, several types of times
corresponding to
several types of measurements have been proposed,  e.g.  the time of 
arrival,  the tunneling time and the time of a quantum clock given by a
phase variable \cite{book}.
In what follows,
we shall consider only the simplest case, the time of arrival 
of a free particle.    

\subsection{Observables and measurements}

In general, the physical meaning of a POVM is given by
the probability postulate:
if an observable $A$ is represented by its spectral measure $E^A(a)$, then
the probability $p_\rho^A (a)$ of getting a result $a$ 
at measuring a quantum state $\rho$ is
\begin{equation}
p_\rho^A (a) = Tr[\rho E^A(a)].   \label{probpost}
\end{equation}

In the  quantum theory of measurement \cite{BLMeasur}, 
the probability postulate (\ref{probpost}) is explained in terms of
a measurement interaction between the measured system and an apparatus 
(a pointer),
which introduces  suitable correlations between them. 
It can be
given in the form of a unitary transformation \cite{BLMeasur}
\begin{equation}
U = e^{i \lambda A \otimes B}, \qquad \lambda = const
                  \label{meastrans}
\end{equation}
which transforms an initial (e.g. factorized pure) state of the system
and the pointer  $\phi \otimes \psi$ into a final state
$$
U |\phi \otimes \psi> = \int_R E^A(da) \phi \otimes e^{i \lambda a B} \psi.
$$
The pointer operator  $B$ 
 must be chosen so that  transformation (\ref{meastrans})
can be interpreted as describing a measurement, i.e.  the final state 
must contain a record of the measurement (eigenvalues $a$ of the 
operator $A$).
In the simplest case, the record  is supposed  to be a shift in the
position $z$ of the pointer particle. Then the corresponding 
operator $B$ in the measurement
interaction can be chosen to be its canonically conjugate 
momentum $\pi_z$ and
the measurement reads \cite{BLMeasur}
($z_0$ denotes the initial position of the pointer)
\begin{equation}
U |\phi \otimes \psi> = \int_R E^A(da) \phi \otimes \psi(z_0 - \lambda a).
                    \label{final}
\end{equation} 
For getting a distinct (classical) record the pointer states
must consist of mutually  disjoint sets and 
after a measurement a possible 
superposition of states must be reduced to a single reading. These 
problems include clarifying the boundary between quantum and classical
world and they are far from being clear;  we don't go into details here
and refer to  e.g. \cite{BLMeasur, Isham}.  

Measurements which
are mathematically described by a POVM and the probability postulate
only, without reference to  details of measuring apparatus, 
are called indirect or ideal measurements.       


\subsection{The Aharonov-Bohm time operator}

An operator corresponding to the time of arrival of a free
particle moving in  one-dimensional space has been proposed
already long ago by Aharonov and Bohm \cite{AB}   
\begin{equation}
{T} = - {m \over 2}( { q}{ p}^{-1} + { p}^{-1}{ q})
  \equiv   - m {1 \over \sqrt {p}} { q}{1 \over \sqrt { p}}\,.
                 \label{ABT}
\end{equation}      
Its eigenvalue problem can be solved in the momentum 
representation \cite{Kijowski, Giannitr,
Egusqu}: 
\begin{equation}
{-im \over \sqrt p }{d \over dp} {1 \over \sqrt p} \Phi_T (p) =
T \Phi_T (p),
\end{equation}
\begin{equation}
\Phi_{T\alpha}(p) = \Theta (\alpha p) \sqrt{{|p|\over m}} 
\exp \bigl({iTp^2 \over 2 m} \bigr) ,
                    \label{ABF}
\end{equation}
where $\alpha = \pm 1$, $T \in( -\infty, \infty)$. 

The Aharonov-Bohm operator (\ref{ABT}) is not self-adjoint and 
its eigenfunctions (\ref{ABF}) are not orthonormal:
$$
 \int_{-\infty}^\infty \Phi^*_{T \alpha}
(p) \Phi_{T^\prime \alpha^\prime} (p) dp =
 \delta_{\alpha \alpha^\prime}
 {1 \over 2}  \bigl( \delta(T- T^\prime) + i P{1 \over 
\pi (T - T^\prime)} \bigr)\,.
$$
The corresponding measurement statistics is given in terms of  the POVM 
\cite{Holevo, BGLOperat, Kijowski, Giannitr}
\begin{equation}
E^T(\Delta T) = \sum_\alpha \int_{\Delta T} \Phi_{T \alpha}^* 
(p^\prime) \Phi_{T \alpha} (p) dT \equiv 
\sum_\alpha \int_{\Delta T} |T \alpha><T \alpha| dT.
                                                      \label{ABPOVM} 
\end{equation}

Let us consider a quantum system in a pure state 
characterized by a density matrix
$\rho (t) = |\phi (t)><\phi (t)|$.
Here $t$ must be understood as a Newtonian time 
parameter which enumerates 
successive states.
According to the probability postulate (\ref{probpost}), 
the probability of arrival during an interval $\Delta T$ reads
\begin{equation}
p^T_\rho (\Delta T;t) = Tr[\rho(t) E^T (\Delta T)] =
\sum_\alpha \int_{\Delta T} |<\phi (t)|T \alpha>|^2 dT.
                             \label{ABstat}
\end{equation}
As in the case of a measurement of  other continuous variables, e.g
the position, the measured quantity here is not an eigenvalue but an 
interval of eigenvalues $\Delta T$. However, as distinct from the 
measurement of the position,  eigenfunctions
$|T \alpha>$ are not only non-normalizable, but also non-orthogonal
 and a general state $|\phi>$
cannot be uniquely given as their linear combination (integral).

Muga et al \cite{Mugaetal} have computed  probability distributions
for normalized and approximately orthogonal Gaussian wavepackets 
(centered at $T$ and with 
width $\delta T$) of Newtonian-time-dependent
eigenfunctions $|T^\prime, \alpha;t> = \exp (-iHt)|T^\prime, \alpha>$. 
They found that the probability distribution of a wave packet 
$\psi(t,x; T, \delta T)$ is peaked 
around the point  $x=0$, $t=T$ and the peaking is   inversely proportional
to the width $\delta T$. Hence 
in this case the time of arrival at $x=0$,  $t=T$,   
can be determined with arbitrarily
sharp accuracy depending of the width of the wave packet.


\subsection{Measurement dynamics}

We may ask if the final state (\ref{final}) of a system 
and an apparatus can also be obtained 
from a suitably constructed Schr\"odinger equation, i.e. if the measurement
can be  considered as a dynamical process given by a suitably constructed 
Hamiltonian \cite{Neumann}. 
Formally, we can  write a two-body Schr\"odinger equation 
for a physical system $(Q,P)$ and a free particle $(z, \pi_z$) 
acting as a pointer  (apparatus) together
with an interaction Hamiltonian which depends explicitly on the
Newtonian time parameter $t$, 
e.g. in the form of an
instantaneous measurement interaction for recording the value of
an observable $A(Q,P)$:    
\begin{equation}
i {\partial \over \partial t} \Psi = H (Q,P,z,\pi_z, t) \Psi,                
                              \label{VNSCH}
\end{equation}
\begin{equation}
H (Q,P,z,\pi_z, t) = H_{sys} (Q,P) + {\pi_z^2 \over 2M} - 
\delta (t) \lambda A(Q,P) \pi_z.
                      \label{VNM}
\end{equation}
A direct calculation confirms that if the pointer is massive (in comparison
with the measured quantum system), i.e. $M \rightarrow \infty$ and the 
pointer dynamics can be neglected,  the solution of the Schr\"odinger
equation (\ref{VNSCH}) is consistent with the final state  (\ref{final}). 

Measurement dynamics (\ref{VNSCH}), (\ref{VNM}) can be  considered also
for the Aharonov-Bohm time operator, $A(P,Q) = T(P,Q)$, 
$H_{sys} (P,Q) = P^2 / 2m$. Detailed calculations have 
been presented for an operational model consisting in  simultaneous 
measurements of
position and momentum in the phase space  \cite{KW, Leonetal}.
The connection between the operational model and 
the Aharonov-Bohm POVM (\ref{ABPOVM}) has been established \cite{Baute}.

However, as indicated by Busch et al \cite{BLMeasur}, 
time dependent Hamiltonian (\ref{VNM})
is not allowed in a mathematically rigorous theory.


\subsection{Direct measurement of time}

For investigating a complete quantum dynamics of a measurement process, 
Aharonov and Bohm \cite{AB} proposed to consider a quantum system 
consisting of three parts: a physical system, an apparatus (a pointer)
and an additional physical particle acting as a clock.
The time of interaction is determined by a physical observable of
the clock particle.
The corresponding three-body Schr\"odinger equation contains
 a   general Hamiltonian which 
includes free Hamiltonians of a physical system $H_{sys}(Q,P)$, 
a pointer $H_a (z, \pi_z)$  and  a clock $H_{cl}(x,p)$
together 
with an interaction Hamiltonian $H_i(Q,P,x,p, z,\pi_z)$:
\begin{equation}
i {\partial \over \partial t} \Psi = H (Q,P, x, p,z,\pi_z) \Psi,              
                              \label{SCH}
\end{equation}
\begin{equation}
H (Q,P,x,p, z,\pi_z) = H_{sys} (Q,P)  + H_a(z,\pi_z) + H_{cl}(x,p)  + 
        H_i(Q,P,x,p, z,\pi_z).
                      \label{HAM}
\end{equation}
The above-mentioned interpretation of three parts of the Hamiltonian, 
$H_{sys}$, $H_a$ and  $H_{cl}$  must follow from the form of the interaction
Hamiltonian. So the specification of $H_i$ is crucial in analyzing the
measurement dynamics.

Aharonov and Bohm \cite{AB} considered a free massive particle as a
clock, $H_{cl} = p^2/2m$. 
They demonstrated that if the wave function of the total system is
factorized, e.g. $\Psi = \psi (x,t)  \otimes \phi (Q,z,t)$ and
the clock state $\psi (x,t)$ is determined from an approximate  
Schr\"odinger equation as a free wave packet, then the general
Hamiltonian (\ref{HAM}) can approximately be reduced to a two-body 
Hamiltonian  (\ref{VNM})
with an interaction Hamiltonian which depends explicitly on the
Newtonian time parameter $t$. They also concluded that
since the time observable belongs to the clock and  
necessarily commutes with any observable 
of the physical system,   
there are no  constraints to the accuracy 
of a record of the time of interaction  by a clock 
at measuring of energy of a physical system. 

Aharonov and Bohm \cite{AB} investigated the
time of arrival  from the point of view of
determining the exact time of measurement of some other physical observable,
e.g. the energy of a quantum system. 
But there can be different 
measurement arrangements which can be interpreted as direct measurements of
the time of arrival itself.  
The first model proposed by Allcock \cite{Allcock}  consisted
of a free (arriving) particle $(x,p)$ and an
interaction Hamiltonian in the form of a complex  
potential $H_i =iV \Theta (x)$, $V= const$ ($\Theta (x)$ denotes the 
 step function).  
The question under investigation was: what is the probability
that the particle in an initial state with support in $x <0$, 
enters the region $x>0$ during a finite time interval $[0,t]$?
The solution $\phi (x,t)$ of the corresponding  Schr\"odinger
equation revealed  a  restriction to the accuracy of the
record of time $t$, $V \delta t \sim 1$ which is not in the form of
the Heisenberg uncertainty relation.      

From the point of view of the general three-body Hamiltonian (\ref{HAM}), 
a direct measurement of  time of arrival 
can be  described by a two-body 
system which consists of a pointer and an arriving free particle acting also
as a clock.  A model whose classical analogue  
(which is more in detail considered in Sec. 3.2)  
gives a distinct record of time of arrival
was presented by Aharonov et al \cite{Aharetal}: 
\begin{equation}
H (x,p,z,\pi_z) = {p^2 \over 2m} +  
 \Theta (-x)\pi_z.        \label{aharetal}
\end{equation}
Here $(x,p)$ is a free particle and $(z, \pi_z)$
is an infinitely massive pointer which records the time of arrival 
of the particle at $x=0$. 
They demonstrated that  the solution of the corresponding 
Schr\"odinger equation 
implies an uncertainty relation between the kinetic energy of the 
particle $E(p)$ and the accuracy of the  pointer which records the
time of arrival,  $\Delta z \simeq \Delta t$,
\begin{equation}
 E(p) \Delta t > 1.  \label{aegenergia}
\end{equation}
This is not a standard quantum mechanical uncertainty because the
corresponding operators belong to different particles and hence 
commute. Analogous restrictions were found also for several improved 
model Hamiltonians \cite{Aharetal}. 

According to conclusions of Muga et al \cite{Mugaetal},
such restrictions are not present in the indirect measurement of the
time of arrival  as given by the Aharonov-Bohm 
POVM (\ref{ABPOVM}). 
Baute et al \cite{Bautesala} argued
that a restriction to the accuracy of an indirect 
measurement of the time of arrival of a free particle  
and its mean energy $<E>$ can be obtained,  
if we consider a nonstandard  uncertainty relation 
introduced by Wigner. More in detail:
the second moment $\tau$  (in respect of an arbitrary reference time $t_0$)
of the probability distribution of the time of arrival  
and its mean energy $<E>$ satisfy 
\begin{equation}
\tau > {\hbar \over <E>}.    \label{Wigner}
\end{equation}   
Since $\tau $ depends on an arbitrary reference
time $t_0$, it  cannot be identified with a usual spread of measurement
results around the mean value and the Wigner uncertainty 
relation (\ref{Wigner}) in general doesn't coincide with the Heisenberg one,
although both quantities, $\tau$ and $<E>$ characterize the same
quantum particle. This makes the relation between conditions
(\ref{aegenergia}) and (\ref{Wigner}) obscure, since the
first one concernes two quantum particles and the second one only one
particle.  

\subsection{Measurements in a closed system}
     
 Hamiltonian (\ref{HAM})
doesn't depend explicitly on the time parameter $t$ and 
a general three-part measurement scheme (\ref{SCH}), (\ref{HAM})
 can be considered as describing a measurement in a closed system
(cf. \cite{Hartle, Engl}).

In \cite{AB}, the clock particle was approximately described by a wave
packet $\psi (x,t)$ and the general
Hamiltonian (\ref{HAM}) was reduced to a two-body Hamiltonian
with an interaction Hamiltonian  depending (via $\psi (x,t)$)  
explicitly on the time parameter $t$, e.g. in a form
of the von Neumann measurement dynamics (\ref{VNM}). 
Casher and Reznik \cite{CR} introduced another approximation 
for determining the time variable $\tau$
by the state of the clock particle 
$ \tau = mx/<p_x>$.
They argued that if the measurement interaction is determined by 
quantized clock time $\tau$, then 
there arises a  constraint on the accuracy $\Delta J$ of a measurement of
an observable $J$  
\begin{equation}
{\Delta J \over J} \geq {\hbar \over (E_{cl} - E_0) \delta T} \, .
\end{equation}
Here $E_{cl}$ is the clock energy which is bounded below by $E_0$ 
(i.e. its Hamiltonian is quadratic in the momentum, such a device
is called  a real clock) 
and $\delta T$ is the duration of the measurement.  
In the von Neumann measurement theory, the only fundamental restriction
to the measurement is the Heisenberg uncertainty relation for 
noncommuting observables and there are no restrictions on the
accuracy of the measurement of  a single observable.
Hence the general three-body quantum measurement
scheme (\ref{SCH}), (\ref{HAM}) contains the von Neumann theory 
in some approximation \cite{AB}, but leads to results which contradict to
it in some other approximations \cite{CR}.  

Aharonov and Reznik  \cite{AR} considered more in detail a measurement of 
the total energy of a system consisting of
a box,  an ideal clock (i.e. a device with the Hamiltonian which is 
linear in the momentum) and a pointer:
\begin{equation}
H = H_{box} + H_{cl} +  
{1 \over 2}[g(\tau) H_{cl} + H_{cl} g(\tau) + 2g(\tau) H_{box}]z \,.   
                          \label{q}
\end{equation}
Here $H_{cl} = -i \hbar \partial / \partial \tau$ is the Hamiltonian of the
clock,  the Hamiltonian
of the pointer $ \pi_z^2 /2M$ vanishes  ($M \rightarrow \infty$) and
$g(\tau)$ is an interaction function normalized as $\int g(\tau) d\tau = 1$.
They demonstrated that the solution of the corresponding Schr\"odinger equation
describes a measurement only if  $g(\tau)$ 
and the pointer coordinate $z$ satisfy a constraint 
\begin{equation}
    g(\tau) z \ll 1 .
\end{equation}
Introducing  an  approximation $g(\tau) \sim 1/\tau_0$ 
($\tau_0$ is the duration of the
measurement)  and taking into account, that 
the accuracy of the measured value 
of the total energy $\Delta E_0$  is given by the record of the 
pointer $\pi_z$, the Heisenberg uncertainty relation for the pointer implies
a constraint 
\begin{equation}
\tau_0 \Delta E_0 \gg \hbar.   \label{22}
\end{equation}
This uncertainty relation  between the duration of the measurement $\tau_0$ 
and accuracy of the measurement record $\Delta E_0$  
is analogous to relation (\ref{aegenergia}) obtained for a model of a
direct measurement of time of arrival and it 
cannot be considered as the Heisenberg uncertainty relation.

We see that at present,
the quantum theory of  the time of arrival
leaves without an answer at least the following  questions:

1. Is its indirect measurement by the Aharonov-Bohm time operator
mathematically and conceptually adequate?

2. If described  as a direct measurement, which is  
the corresponding Schr\"odinger equation?

3. Which is the status of  non-Heisenberg uncertainty relations
between time and energy in direct and indirect measurements?

We argue that some of these  questions  arise already 
in the classical theory of measurement
of time.

\section{The classical theory}
\subsection{External time and internal time}

Classical canonical (Hamiltonian) mechanics is formulated in a phase
space $(p,q)$ and the dynamics is determined by a Hamiltonian
$H(p,q,t)$. Numerically, the Hamiltonian is equal to the total energy $E$
of the physical system  under consideration. Equations of motion
can be derived from the canonical integral
\begin{equation}
S = \int(p{\dot q} - H(p,q,t))dt\,, \quad {\dot q}\equiv {dq(t) \over dt}
\end{equation}
as the Euler--Lagrange equations. Time $t$ is a 
parameter along trajectories, 
$p=p(t), q=q(t)$.

Observable properties are represented by canonical coordinates 
and their functions. For introducing an observable of time, 
$t$ must be on an equal footing  with space coordinates
$q$. There are two possible ways for achieving this.

1. In the parametrized form of Hamiltonian dynamics \cite{Lanczos}, 
the parameter $t$ is considered as an additional canonical time coordinate.
The corresponding canonically conjugate momentum can be shown to
be the total energy, $p_t = -H$, and the canonical integral takes a 
symmetric form
\begin{equation}
S= \int (p{\dot q} + p_t{\dot t})d\tau\,.
                    \label{tpkan}
\end{equation}
The Euler--Lagrange equations are parametrized with an arbitrary 
parameter $\tau$
and hold on a constraint surface    
 $p_t+H(p,q,t)=0 $. 
Alternatively, we can  take the totally parametrized
canonical integral (\ref{tpkan})
as a starting point, then the physics is determined by the 
constraint equation
which must be added. If we add the conventional constraint equation
 which is linear in $p_t$, we get the conventional
mechanics (energy mechanics). But in general we are 
free to choose the constraint equation as
we please. In this way we can get unconventional mechanics, 
interpretation of which
must be deduced from the equations. An example of an unconventional 
mechanics is so-called time mechanics \cite{KP, KK}, which follows from
a constraint equation which is linear in $t$.

2. In the internal time approach by Rovelli \cite{Rov1},
the parameter $t$ is eliminated from the equations of motion and 
trajectories are parametrized by one
of the canonical coordinates, e.g by $q_1$, which describes  
the position of a physical clock particle. 
The procedure is in general 
possible if the Hamiltonian doesn't depend on time $t$ explicitly.
Then it is equivalent to the standard procedure of lowering of the 
dimension of the phase space using the Hamiltonian as the first integral,  
  $H(p_i,q_i) = h = const$ \cite{Arnold}. 
Let us solve the latter relation in respect of  e.g. $p_1$:
$$ p_1 = K(p_j, q_j, q_1, h), \qquad j \not= 1.
$$  
According to the Arnol'd theorem \cite{Arnold}, 
on a surface of constant energy, $H(p_i,q_i) = h$, 
the trajectories can be parametrized by  the corresponding coordinate
$q_1$ and their equations are
\begin{equation}
{dp_j \over dq_1 }= {\partial K \over \partial q_j}, \qquad
{dq_j \over dq_1 } = -{\partial K \over \partial p_j}.
                              \label{K}
\end{equation} 

As an example, let us consider  4-dimensional phase space  $(P_x,x, p_t,t)$. 
If we introduce the usual energy constraint
$p_t + H_1(P_x,x,t) =0$, we get the conventional canonical equations 
for a particle in  one-dimensional $x$-space, whose trajectories are 
parametrized by time $t$  
\begin{equation}  
{dP_x \over dt }=- {\partial H_1 \over \partial x}, \qquad
{dx \over dt } = {\partial H_1 \over \partial P_x}.
                                     \label{H}
\end{equation}
Alternatively, we can  consider the constraint as a general Hamiltonian 
$H_2(P_x,x,p_t,t)$  which
doesn't depend explicitly on the external time parameter $\tau$ and use
the Arnol'd theorem for lowering the dimension of the phase space.
If $H_2$ is linear in the momentum $p_t$, i.e. $H_2 = p_t + H_1$, 
then equations of motion
in  $x$-space (\ref{K}) and (\ref{H}) coincide (the latter 
statement is evidently
true also in a general $2n$-dimensional case). 

But there is a difference in the interpretation.
In the first case, we are considering the dynamics of a particle with
the Hamiltonian  $H_1(P_x,x,t)$.
In the latter case,  both canonical coordinates 
$x,t$ are at first considered as possible trajectories of 
physical particles, positions of which are measurable. The particle
which can move in the $t$-space only is a physical clock and the particle
which can move in the $x$-space only is a physical system under 
investigation.         
Let us interprete the total Hamiltonian $H_2(P_x,x,p_t,t)$
as a sum of corresponding free Hamiltonians
plus an interaction term. Then the usual energy constraint 
$p_t + H_1(P_x,x,t) =H_2$  describes a  
clock particle with a linear free Hamiltonian $H_{cl}= p_t$. 
It determines the trajectory of a free 
clock particle as $p_t = const$, $ t=\tau + \tau_0$, 
where $\tau$ is a parameter
along the trajectory. Such a device is known as  an ideal clock. 
It is different from 
a real clock described by a free Hamiltonian which is quadratic in its
momentum, $H_{cl} =p_t^2 /2M$ and the trajectory of which reads
$p_t = p_t^0 = const$, $t = p_t^0 \tau + \tau_0$. 

Finally, let us note that
since a trajectory $(p(t), q(t))$ is a map $R \rightarrow R^{2n}$,
i.e. canonical coordinates and the time parameter are both real numbers, 
equations of type $x=t$ are mathematically meaningful.   
 
\subsection{Classical measurement of time}

The measurement of Newtonian time $t$ can be realized as a measurement
of the position of a pointer particle with mass $M$ which is moving freely
with a constant momentum $P_y^0$:
\begin{equation}
y(t) = {P_y^0 \over M }t + y(0).
\end{equation}
We can achieve also $y=t$ by choosing suitable values of constants.

Let us consider a physical system given by a Hamiltonian
$H_0 (p,q)$  and let $A(p,q)$ be
a variable we want to measure at a time $t_0$. 
Let us describe the measurement dynamics by the same Hamiltonian (\ref{VNM})
as in the quantum theory  \cite{Peres} 
\begin{equation}
H = H_0(p,q) + {P_y^2 \over 2M} + \delta(t-t_0) A(p,q) P_y.
                                 \label{peres1}
\end{equation}
The trajectory of the pointer is  
\begin{equation} 
y(t)= {P_y^0 \over M}t+ y(0) + A(t_0) .
                    \label{osuti}
\end{equation}     
We see that now the pointer is recording two distinct physical quantities:
the Newtonian time $t$ (as in the previous example) and the value
$A(t_0)$. The latter can be read out as a constant shift in the
position of the pointer at $t>t_0$ in comparison with the unperturbed
trajectory at $t< t_0$. We see that the measurement of a quantity $A$ at a 
time $t_0$ is a nonlocal procedure in  time $t$ in the sense that  
at first we
must determine the undisturbed trajectory of the pointer and only then we can
read out the shift proportional to the measured quantity. Usually the 
nonlocality in time is eliminated by an assumption that the mass of the
pointer is very big and in the limit of an infinite mass 
($M\rightarrow \infty$) the position of the pointer at $t>t_0$
 records only a constant shift $y(t) - y(t_0) = A(t_0)$. 

Let the physical system under consideration be a freely moving particle
with Hamiltonian $H_0 (P_x, x) = (P_x)^2 /2m$. 
The property we want
to measure is its time of arrival  from initial position $x_0= x(0)$ to  point
$x$, i.e. $A (P_x, x)  =  (x-x_0)m /P_x$. 
The trajectory of the pointer reads
\begin{equation} 
y(t)= {P_y^0 \over M}t+y(0)+  (x(t_0)-x_0){m \over P_x}
= {P_y^0 \over M}t  + y(0) +t_0.  \label{yt} 
\end{equation}     
But in such a measurement arrangement, the time $t_0$ has a double meaning: 
it is the time of the measurement 
as prescribed by the Hamiltonian (\ref{peres1}) and the  quantity we want 
to measure. 
The time of switching on the measurement interaction $t_0$ equals to the 
measured quantity and cannot be prescribed arbitrarily.    

An interaction  Hamiltonian which doesn't fix the time of the measurement
$t_0$ in advance was  proposed by Aharonov et al 
\cite{Aharetal} 
\begin{equation} 
H(P_x, x; P_y, y) = {P_x^2 \over 2m} + {P_y^2 \over 2M} + \Theta(-x) P_y.
                          \label{theta}
\end{equation}
The equations of motion of the pointer are  
\begin{equation}
{\dot P_y} = 0, \qquad
{\dot y} = {P_y \over M} + \Theta(-x)
\end{equation}
and its trajectory  is 
\begin{equation}
y(t) = {P_y^0 \over M}t + y(0) + \int_0^t \Theta(-x(\tau))d\tau,
 \qquad  P_y =P_y^0=const.
\end{equation}
 The pointer is moving from its initial
position $y(0)$ with a constant velocity ${\dot y} =P_y^0 / M +1$
till the arrival of the particle at $x=0$ and after that begins to move
undisturbed.  If the pointer is infinitely massive, 
it moves with a unit velocity
${\dot y} =1$ and stops at the moment $t_0$ 
when the particle  arrives 
at point $x=0$ from its initial position $x(0) = x_0 <0$.
The situation is different in comparison with that considered in the previous
example (\ref{yt}):
the interaction between the  pointer and the particle 
takes place before the measurement, not just at the time of measurement
$t_0$. This means that here we don't
measure the time of arrival of a freely moving particle but of a 
particle in an interaction with the pointer. The 
backreaction of the interaction to the particle 
can be seen from the conservation of energy
\begin{equation}
   {(P_x (x))^2\over 2m} + {(P_y^0)^2 \over 2M} + \Theta(-x) P_y^0 = E= const.
\end{equation}
Aharonov et al \cite{Aharetal} argue that we can take $P_y^0=0$,
then the particle moves freely and the motion of the pointer consists 
only of a constant shift $t_0$. But such a pointer has neither 
kinetic nor potential energy and clearly is not a generic case.

\subsection{A general model for a real clock and a pointer}
 
Let us now consider the Hamiltonian (\ref{theta}) with a general
interaction function $g(x)$ instead of $\Theta(-x)$:
\begin{equation} 
H(P_x, x; P_y, y) = {P_x^2 \over 2m} + {P_y^2 \over 2M} + g(x) P_y.
                   \label{gx}
\end{equation}
The  equations of motion generated by (\ref{gx}) read 
\begin{equation}
{\dot P_y} =0, \qquad {\dot y} = {P_y \over M} + g(x),
\end{equation}
\begin{equation}
{\dot P_x} = { dg \over dx} P_y , \qquad {\dot x} = {P_x \over m}.
\end{equation}
For $x(t)$ we get the following equation
\begin{equation}
{\ddot x}=-{P^0_y \over m} {dg \over dx}, \qquad  P_y^0 =const = P_y.
\end{equation}
Its general solution  establishes a
relation between internal clock time $x$ and external time $t$
\begin{equation}
C_2 \pm t = \int {dx \over \sqrt{C_1 - 2P_y^0 g(x)/m}}.
\end{equation}
Here $C_1$ and  $C_2$ are constants of integration.

The  equation of motion of the pointer in respect to the internal time
$x$ now reads:
\begin{equation}
{ {\dot y}\over {\dot x}} \equiv {dy \over dx} =
{ P_y^0 /M + g(x) \over \pm \sqrt{ C_1 - 2 P_y^0 g(x)/m}}.
\end{equation}

In the case of an infinitely massive pointer ($M \rightarrow \infty$)
the first term in the numerator vanishes, but a dependence on the
pointer momentum $P_y^0$ remains in the denominator. A pointer with
an exactly vanishing momentum is an additional and unrealistic condition.
The influence of a small but nonvanishing $P_y^0$ is a multiplicative 
error in the shift of the pointer position
\begin{equation}
{dy \over dx} = { g(x) \over \pm \sqrt{ C_1}} (1 +{ P_y^0 g(x)\over mC_1}+
 \ldots ).
\end{equation} 

If $P_y^0 = 0$  then $P_x=P_x^0=const$. Now the constant of integration 
$C_1$ can be given in terms of  $P_x^0$ as
$\pm \sqrt{C_1}=  P_x^0/m$ and the position of the
pointer is
$$
y\Big|_{P_y^0=0} = {m \over P_x^0} \int g(x)dx +C.
$$ 

In the case of the interaction considered by Aharonov et al \cite{Aharetal},
 $g(x) = \Theta (-x)$, the equation of the pointer can be integrated as 
\begin{equation}
y =\int^\infty_{x_0} {m \over P_x^0} \Theta (-x)dx =
 {m \over P_x^0}(-x_0) = t_0.
\end{equation}
The pointer stops and indicates the external 
time $t_0$ necessary
for the free particle with constant momentum $P_x^0$ to move 
from initial position $x_0< 0, t=0$ to the detector at point $x=0$.

However, this "good measurement" takes place only if $P_y^0 = 0$. If
$P_y^0 \equiv \Delta P_y^0 \not= 0$ then 
the condition for an "approximately good measurement"
reads
\begin{equation}
{g(x) \Delta P_y^0 \over mC_1} \ll 1.
\end{equation}  

The measurement arrangement (\ref{peres1}), (\ref{osuti}) of an observable
$A(p,q)$ can also be reformulated in terms of the internal time $x$:
\begin{equation}
H = H_0(p,q) + {P_y^2 \over 2M} +  {P_x^2 \over 2m}+ g(x) A(p,q) P_y.
\end{equation}
The equation of motion of an infinitely massive pointer 
($M \rightarrow \infty$) reads
\begin{equation}
{dy \over dx }= {g(x) A \over  \pm \sqrt{C - 2A P_y^0 g(x)/ m}}
\end{equation}
where $C$ is a constant of integration. The condition for a 
"good measurement" reads
\begin{equation}
{A P_y^0 g(x)\over mC} \ll 1 \,.    \label{gm}
\end{equation}
If we approximate  $g(x) =  \delta(x) \sim 1/ x_0$, where $x_0$ is the
duration of the measurement interaction in the internal time and 
take into account that in the best case $P_y^0 =0+\Delta P_y^0$ and  
an error $\Delta y$ in the pointer position is proportional to the
measurement error $\Delta A$, $\Delta y = \Delta A / \sqrt{C}$,
then from  the condition for a "good measurement" (\ref{gm}) it follows that
\begin{equation}
\Delta y \Delta P_y^0 \ll \sqrt{C} m x_0  {\Delta A \over A}\,.
                  \label{clgm}
\end {equation}
We know that in the quantum theory the left hand side has a lower limit
$\hbar \ll  \Delta y \Delta P_y^0$. In the classical theory, it has no 
absolute lower limit, but still its vanishing is unrealistic.  
 Note that the first factor on the right hand
side depends on the clock particle and the second one on the measured
system. 

\subsection{Measurement of total energy in internal time}

In Sec. 2.6, we considered a quantum measurement of total  energy  
of a closed system with an ideal clock as investigated  by 
Aharonov and Reznik \cite{AR}. Let us now consider the corresponding
classical model given by  the same  Hamiltonian (\ref{q})
\begin{equation}
H = H_{box} + P_x  + {P_z^2 \over 2M} + 
g(x)(H_{box}+ P_x + {P_z^2 \over 2M})z.
\end{equation}
We can take the pointer to be infinitely massive, $M\rightarrow \infty$,
and ignore its kinetic energy terms. The measurement outcome is recorded
by a change in the momentum of the pointer, its position remains
constant, $z=z_0$.   
Let the energy of the box be 
a constant of motion, $H_{box} = H^0$. Now the following equation of
motion for the pointer momentum $P_z$ in respect of the internal time $x$
can be derived
\begin{equation}
{dP_z \over dx } =- {g(x) ( H^0+C) \over (1+g(x)z_0)^2}, \qquad C=H-H^0,
                   \label{linkell}
\end{equation}
where $C$ is a constant of integration. The condition 
for a "good measurement" is
\begin{equation}
g(x)z_0 \ll 1.        \label{clgmt}
\end{equation}
If we take the interaction function to be inversely proportional to the
duration of the measurement, $g \sim 1/x_0$, and the pointer position is 
$z_0 = 0+\Delta z$, then condition (\ref{clgmt}) reads
\begin{equation}
\Delta z \ll x_0.
\end{equation}   
But since $P_z$ records the measured energy $H$, the measurement
error is $\Delta P_z = \Delta H$. As a result we get a relation
between the duration of the measurement in respect of the internal 
clock time and measured total energy:
\begin{equation}
\Delta z \Delta P_z \ll x_0 H.
\end{equation}
This can be considered as a classical analogue of the quantum
relation  (\ref{22}). If there is a
lower limit for the l.h.s., then there is a lower limit for the
product of measurement duration in internal clock time and
measured total energy of the system.

The problem can mathematically be solved also in the case of a 
real clock with a quadratic
 Hamiltonian  $H_{cl}=P^2_x/2m$. The equation of
motion of the pointer in respect of the internal time showed by the
real clock is
\begin{equation}
{dP_z \over dx } =- {g(x) \over (1+gz_0)^{3/2}} {C +2mH^0 \over
2\sqrt{C - 2mH^0 z_0 g}}, \qquad C=2m(H-H^0),
                   \label{realkell}
\end{equation}
where $C$ is a constant of integration. The condition  
of a "good measurement" (\ref{clgmt}) must now be supplemented with
a second condition 
$ \sqrt{C - 2m H^0z_0 g} \sim m $,
from which  a condition for the clock $P_x \sim m$ follows.
It amounts to an assumption that the clock particle 
must move with a unit velocity,  $dx/dt =1$. 
We see that in this case the model with a real clock doesn't 
add anything essentially new in comparison with the model
with an ideal clock.

\section{Conclusions and discussion}

It seems that the problem of  time has its beginning
in the classical mechanics. If we use the Hamiltonian formalism and agree
that the only measurable quantities are canonical coordinates (and
canonical momenta), then there are two distinct possibilities for
introducing the measurement of time: either we allow  direct
measurements of time coordinate $t$ in an extended phase space, 
or we visualize the time
flow by a change in the position  $x$ of a  clock 
particle given in an unextended phase space. 

As a result, there are two distinct ways for introducing the
notion of a nontrivial quantum mechanical time. 

1. In the parametrized Hamiltonian dynamics the energy and time
are canonically conjugate coordinates and in a  canonical 
quantization procedure they are replaced by a pair of canonically
conjugate operators. Their dependence on the other canonical 
operators follows from the constraint equation, e.g. 
$p_t+H(p,q,t)=0 $ determines  the energy operator 
$E(p,q,t) \equiv -p_t= H(p,q,t) $
and the corresponding time operator $T(p,q)$ must be found from the
canonical commutation relation $[T,H]=-iI$. 
(Or alternatively, a constraint equation
linear in $t$, $t+T(p,q,p_t)=0 $, determines the time operator and the
energy operator must be found from the canonical commutation relation, see
\cite{KP,KK}.)

2. In  lowering of  dimension of the phase space using 
the Hamiltonian as the first integral, the resulting trajectories
are parametrized by a canonical coordinate $q_1$ which corresponds to
the internal time recorded by the position of a clock particle.
In the canonical quantization procedure  we identify the canonical 
position operator $q_1$   with the internal
time operator. 
Now we can perform general investigations  in the Heisenberg  
representation \cite{Rov1}, but this has given us only very 
general insights.
If we use the Schr\"odinger representation,  
we can identify eigenvalues of the position operator $q_1$ 
with the time parameter which occurs in the Schr\"odinger
equation,  $q_1 = t$,  
and determine probabilities from the total wave function 
 \cite{Aharetal, CR}.
Here the results depend crucially on our choice of the Hamiltonian
in the corresponding Schr\"odinger equation.    

These two possibilities for introducing a  quantum
mechanical time  differ in several aspects.
In the  first case, the time is an operator canonically conjugate 
to the energy,
in the second case, it is a position operator canonically conjugate to
a momentum operator.  In the first case, the classical counterpart 
of the operator
algebra is an extended phase space where canonical
variables $(t, p_t)$ are assumed to describe an ideal
clock, in the second case, it is an unextended phase space where 
a pair of canonical variables $(q,p)$ are assigned to a real clock particle.
The distinction in the interpretation of clock variables results in
the choice of clock Hamiltonians: in the first case the usual
constraint equation entails a Hamiltonian which is linear in the
momentum $p_t$ (an ideal clock), in the second case it is natural
to choose a Hamiltonian which is quadratic in the clock momentum $p$
(a real clock).   

Is it possible to find a correct Hamiltonian and a well-defined
measurement interaction  for describing a quantum
measurement of  time of arrival? The corresponding classical theory
gives us the following hints.

1. It is not reasonable to apply the conventional measurement theory
with an instantaneous measurement interaction in the case of a 
measurement of the time of arrival. However, it is less problematic 
in the quantum theory,
where the measured observable is given by the Aharonov-Bohm time 
operator and the measurement
interaction depends on the time parameter; in the classical theory both 
types of time coincide and this leads to the above-given statement.   

2. It is not clear how to build an ideal clock with a Hamiltonian
which is linear in the momentum.  

3. If we use a real clock particle with a quadratic Hamiltonian, 
its interaction with a pointer particle gives rise
to  measurement records which are reasonable only if certain
"good measurement " conditions hold (e.g. (\ref{clgm}) or (\ref{clgmt})).

\bigskip
{\bf Acknowledgements}
\medskip

The authors are grateful to Kaupo Palo and Paavo Helde for discussions
and to J.G.~Muga for sending us the comprehensive review \cite{physrep}.
This work has been supported by the Estonian Science Foundation
via Grant No 3870.    


\end{document}